\documentclass[runningheads]{llncs}
\usepackage[english]{babel}

\usepackage{adjustbox}
\usepackage{graphicx}
\graphicspath{{img/}} 
\usepackage[caption=false]{subfig}
\usepackage{tabularx}
\usepackage{hyperref}
\usepackage{bm}        
\usepackage{amsmath,amssymb,}
\usepackage{hyperref}
\usepackage{color}
\usepackage[dvipsnames]{xcolor}
\definecolor{purple}{rgb}{0.58,0.0,0.83}
\usepackage{caption}
\usepackage{tabularx}
\usepackage{schemata}
\usepackage{tikz}
\usepackage[qm]{qcircuit}
\usepackage{caption}
\usepackage{ragged2e}
\usepackage{mathtools}

\DeclarePairedDelimiter\ket{\lvert}{\rangle}
\DeclarePairedDelimiterX\braket[2]{\langle}{\rangle}{#1\,\delimsize\vert\,\mathopen{}#2}

\usetikzlibrary{backgrounds}
\usetikzlibrary{shadows,fit,shapes,chains,backgrounds,scopes}
\usetikzlibrary{positioning,arrows} 
\usetikzlibrary{patterns}
\usepackage{eso-pic}
\usetikzlibrary{babel}
\usetikzlibrary{patterns}
\usetikzlibrary{decorations.pathreplacing}
\usetikzlibrary{calc,trees,shapes.geometric,decorations.pathreplacing,arrows.meta,decorations.pathmorphing,shapes,matrix,shapes.symbols}

\definecolor{azulipn}{RGB}{174,182,211}
\definecolor{ipn-guinda}{RGB}{90,18,54}
\definecolor{ipn-blanco}{RGB}{241,241,241}
\definecolor{goodDarkGreen}{RGB}{42	,109,60}
\definecolor{goodDarkOrange}{RGB}{164,101,10}
\definecolor{goodDarkRed}{RGB}{147,10,10}
\usepackage{orcidlink}

\begin{document}
\title{Quantum Block-Matching Algorithm using Dissimilarity Measure\thanks{CONAHCYT, IPN and qbraid.}}
%
%
\author{M. Martínez-Felipe\inst{1}\orcidlink{0000-0003-4806-1390} \and
J. Montiel-Pérez\inst{1}\orcidlink{0000-0002-0214-1080} \and
Victor Onofre\inst{2}\orcidlink{0000-0002-6175-9171} \and
A. Maldonado-Romo\inst{1,2,3}\orcidlink{0000-0002-6342-3508} \and
Ricky Young \inst{2,3}
}
\authorrunning{M. Martínez-Felipe et al.}
\institute{Centro de Investigaci\'on en Computaci\'on, Instituto Polit\'ecnico Nacional,  07738, Ciudad de M\'exico, M\'exico  \email{\{mmartinezf2020,jyalja,amaldonador2021\}@cic.ipn.mx}\and
Quantum Open Source Foundation \\ 
\email{\{vonofre68\}@gmail.com} \\ \url{https://www.qosf.org/}
 \and
qBraid, 60615, Chicago, United States
\\ \email{\{rickyyoung\}@qbraid.com}}

\maketitle              
\begin{abstract}

Finding groups of similar image blocks within an ample search area is often necessary in different applications, such as video compression, image clustering, vector quantization, and nonlocal noise reduction. A block-matching algorithm that uses a dissimilarity measure can be applied in such scenarios.  
In this work, a measure that utilizes the quantum Fourier transform or the Swap test based on the Euclidean distance is proposed. Experiments on small cases with ideal and noisy simulations are implemented. In the case of the Swap test, the IBM and IonQ quantum devices have been used, demonstrating potential for future near-term applications.

\end{abstract}
\section{Introduction}
\begin{figure}[h]
    \centering
    \includegraphics[width=.7\linewidth]{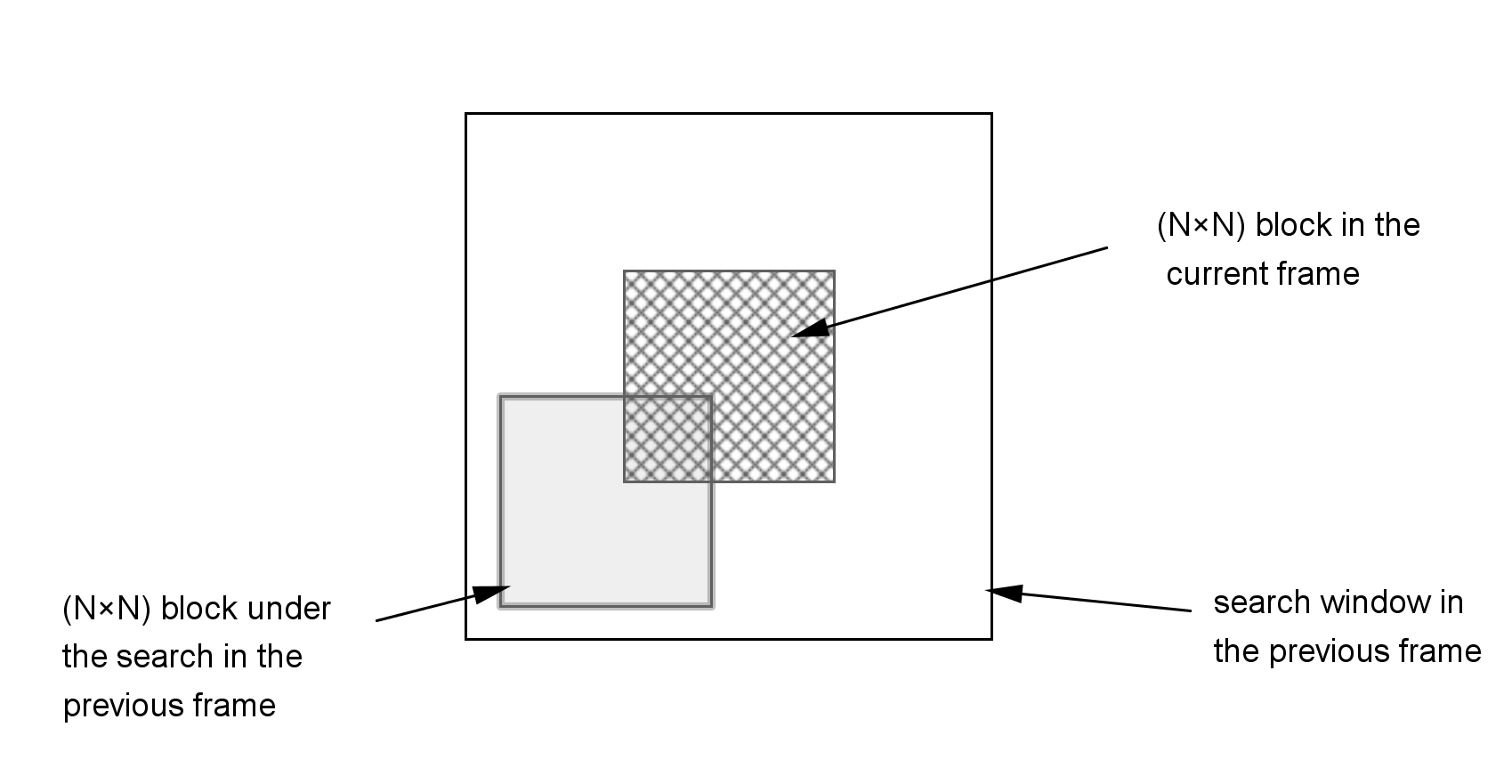}
    \vspace{1ex}
  \caption{A representation of block-matching algorithms.}
  \label{fig:image_1}
\end{figure}

Block-matching (BM) is a way to locate patch blocks in a sequence of digital images; This BM algorithms use a similarity or dissimilarity measure to compare images or some regions of an image \cite{01Libro}. A representation of BM is shown in \autoref{fig:image_1}. There are various applications for BM, mainly in video compression, image clustering, nonlocal noise reduction, and vector quantization \cite{Felipe20193169}.

Quantum image processing (QIP) opens the possibility of design, store, develop, and implement new methods applied to image processing algorithms. For example, there are some quantum image representations such as flexible representation (FRQI) \cite{le2011flexible}, multi-channel (MCQI) \cite{sun2013rgb}, and novel enhanced (NEQR) \cite{zhang2013neqr}, which is one of the earlier forms of quantum image representation. It uses a normalized superposition to store pixels in an image. NEQR was created to leverage the basis state of a qubit sequence to store the image's grayscale value.

Developing new methods and algorithms using the quantum computing framework offers a possible advantage over classical computing. Nevertheless, these quantum computers contain hundreds of noisy qubits and perform imperfect operations in a limited coherence time. We are in the Noisy intermediate-scale quantum (NISQ) era \cite{preskill2018quantum}, it is necessary to mitigate the error rates. Recent theoretical works have shown improvements in implementing error mitigation techniques with these NISQ devices \cite{01NISQ,02NISQ,05NISQ}, opening opportunities for near-term applications.


In this work, we develop quantum circuits in order to map the BM problem in a noisy environment using dissimilarity measurement based on Euclidean distance. Since low-complexity block-matching algorithms based on patch recognition in image processing are required for comparing images, exploring some regions of one image plays a crucial role in standard video codec due mainly to the motion estimation process but also in applications previously mentioned. Nevertheless, using the properties of quantum computing as superposition or the amplitude encoding of a qubit, the quantum block-matching algorithm could help to reduce the complexity in many operations since $n = log_{2}(M) + 1$ we could use only $n$ qubits to encode $M$ classical data. 

This paper is structured as follows: \autoref{Sect:2} explores the block-matching algorithms, and \autoref{Sect:3} explains the methodology used. \autoref{Sect:4} is dedicated to explaining the result for different quantum noise models. Finally, \autoref{Sect:5} summarizes this proposal work and experimental results.

\section{Block-matching algorithms} \label{Sect:2}

In previous works \cite{24Springer}, there has been an implementation of the generalized quantum image representation (GQIR), where the size of an image is $2^m \times 2^m$ and $I \left(x, y\right) \in \{0, 1, \cdots, 2^{n}-1\},$ where $m$ is an integer related to the size of images and $n$ the number of qubits. The idea is to find a small image in a big image using Grover's algorithm. Nevertheless, noisy environments to find a particular patch in an image are not applied. Although this algorithm improved in \cite{26Springer} to solve the error of matching at the upper left corner of one pixel, maintaining an advantage with less complexity than classical algorithms. On the other hand, there are methods based on the quantum-classical approach. In \cite{quantumHybrid}, a method for image matching with a scale-invariant feature transform (SIFT) algorithm is presented. In this last work, feature extraction from images is done to assign an orientation to each key point location. Although these related works present novel solutions, none of the previous works study how quantum noise affects the results in a Quantum Processing Unit (QPU). These works are compared in the \autoref{tab1}.

\begin{table}
\caption{Comparison between block-matching related works.}\label{tab1}
\begin{tabular}{|p{.45\textwidth}|p{.35\textwidth}|p{.15\textwidth}|}
\hline
\textbf{Related works}        & \textbf{Algorithms} & \textbf{Noisy models } \\ 
\hline 
Analysis and improvement of the quantum image
matching \cite{26Springer}.   &  NEQR + Grover’s algorithm  + improvement                 & No \\ 
Quantum image matching \cite{24Springer}. & NEQR + Grover’s algorithm                  & No                     \\  
A Hybrid Quantum Image-Matching Algorithm  \cite{quantumHybrid}. & SIFT + amplitude encoding    + similarity measure               & No\\ 
Proposal methodology & Hierarchical search + encode + Quantum Swap Test or Quantum Fourier Transform & Yes\\
\hline
\end{tabular}
\end{table}

\subsection{QFT approach} \label{Sect:2.1}

The Draper adder \cite{draper2000addition}, is arguably one of the most elegant quantum adders, as it directly invokes quantum properties to perform addition. The insight behind the algorithm is that the Fourier transform can be used to translate phase shifts into a bit shift. It follows the implementation with an adder by applying a Fourier transform, applying appropriate phase shifts, and then the Fourier transform inverse is done. Unlike many other adders that have been proposed, the Draper adder does not have a natural classical counterpart.

\subsection{Swap Test} \label{Sect:2.2}

\begin{figure}
\centering
     \begin{tikzpicture}
\node[] (circuit){\Qcircuit @C=1em @R=1em {
\lstick{\ket{0}}& \gate{H} & \qw   & \qw & \ctrl{2} & \gate{H} & \meter \\
\lstick{q_{1}} & \qw & \gate{\psi}  & \qw & \qswap & \qw & \qw \\
\lstick{q_{2}} & \qw & \multigate{2}{\phi} &  \qw & \qswap & \qw & \qw \\
\lstick{q_{3}} & \qw & \ghost{\phi} & \qw & \qw & \qw & \qw\\
\lstick{q_{4}} & \qw & \ghost{\phi} & \qw & \qw & \qw & \qw\\
\lstick{C} & \qw & \qw &\qw &\qw & \qw & \qw \cwx[-5]}};
    \end{tikzpicture}
    \caption{Quantum circuit swap test to compute the quantum Euclidean distance.}
    \label{fig:QED}
\end{figure}
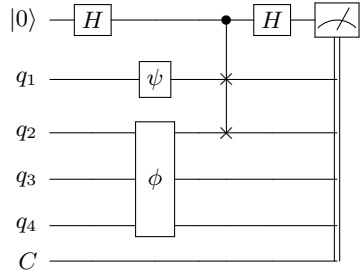

A dissimilarity measure based on a swap test \cite{01QIEEE} to compare two states, $\ket{\phi}$ and $\ket{\psi}$ to compute the Euclidean distance \cite{01Phys,01NeuralNetworks} is implemented. First, the classical data, represented by the vectors $A$ nd $B$ is encoded in quantum states:

\begin{equation}\label{eqn:1}
\begin{split}
    A \longrightarrow \ket{A} = \frac{1}{|A|} \sum_{i}A_{i} \ket{q_{i}} \\
    B \longrightarrow \ket{B} = \frac{1}{|B|} \sum_{i}B_{i} \ket{q_{i}}
\end{split}
\end{equation}

\noindent then, the quantum states $\ket{\psi}$ and $\ket{\phi}$ are defined:

\begin{equation}\label{eqn:2}
\begin{split}
    \ket{\psi} = \frac{\ket{0} \otimes \ket{A} + \ket{1} \otimes \ket{B}}{\sqrt{2}}\\
    \ket{\psi} = \frac{|A|\ket{0}- |B|\ket{1}}{\sqrt{Z}}
\end{split}
\end{equation}

\noindent where, $Z= |A|^{2}+|B|^{2}$. The advantage of this methodology is to perform a few qubits, specifically $n = \log_{2}(M) + 1$ where $n$ is the number of qubits and $M$ the classical data coded with amplitude embedding \cite{01QBook}. The quantum circuit is shown in the \autoref{fig:QED}. Now, to obtain the quantum Euclidean distance, the following equation is solved:

\begin{equation}\label{eqn:3}
    D^{2} = 2Z|\braket{\phi}{\psi}|^2
\end{equation}

\section{Methodology} \label{Sect:3}

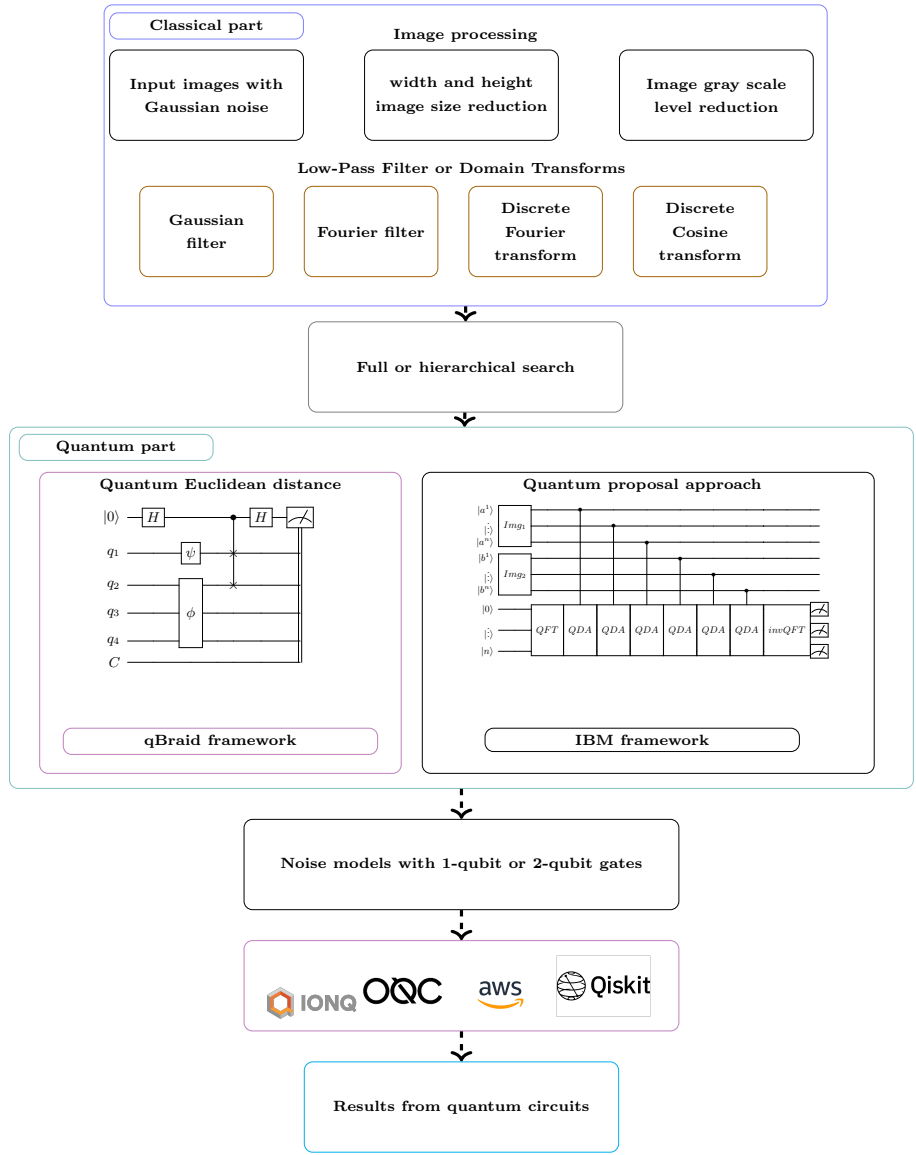
\begin{figure}
\resizebox{\textwidth}{!} {
\begin{tikzpicture} 
   \tikzstyle{rec}=[align=center,rectangle,rounded corners,minimum size=3cm]

\node  (A2) [draw=blue!50,align=center,rectangle,rounded corners,minimum height=5cm,minimum 
width=12cm, text width=5.5cm] at (0.3,3.5)  {\tiny};

\node (A3) [align=center, text width=3 cm,draw=blue!50,rounded corners] at (-4,5.65) { \textbf{\scriptsize Classical part}};

\node (E1) [align=center, text width=3 cm] at (.3,5.5) { \textbf{\scriptsize Image processing}};

\node  (B) [align=center,rectangle,draw=black,rounded corners,minimum height=1.5cm,minimum width=3cm, text width=3cm] at (-4,4.5) { \textbf{\scriptsize Input images with Gaussian noise}};

\node  (C) [right = 1cm of B,align=center,rectangle,draw=black,rounded corners,minimum height=1.5cm,minimum width=3cm, text width=3cm] {\textbf{\scriptsize width and height image size reduction}};

\node  (D) [right = 1cm of C,align=center,rectangle,draw=black,rounded corners,minimum height=1.5cm,minimum width=3cm, text width=3cm] {\textbf{\scriptsize Image gray scale level reduction}};


\node (E1) [below = .25cm of C,align=center, text width=6 cm] { \textbf{\scriptsize Low-Pass Filter or Domain Transforms}};

\node  (F) [below = .75 cm of B,align=center,rectangle,draw=goodDarkOrange,rounded corners,minimum height=1.5cm,minimum width=2cm, text width=2cm] {\textbf{\scriptsize Gaussian filter}};

\node  (G) [right = .5 cm of F,align=center,rectangle,draw=goodDarkOrange,rounded corners,minimum height=1.5cm,minimum width=2cm, text width=2cm] {\textbf{\scriptsize Fourier filter}};

\node  (H) [right = .5 cm of G,align=center,rectangle,draw=goodDarkOrange,rounded corners,minimum height=1.5cm,minimum width=2cm, text width=2cm] {\textbf{\scriptsize Discrete Fourier transform}};

\node  (I) [right = .5 cm of H,align=center,rectangle,draw=goodDarkOrange,rounded corners,minimum height=1.5cm,minimum width=2cm, text width=2cm] {\textbf{\scriptsize Discrete Cosine transform}};


\node  (J) [below = .25cm of A2,align=center,rectangle,draw=gray,rounded corners,minimum height=1.5cm,minimum width=5cm, text width=5cm] {\textbf{\scriptsize Full or hierarchical search}};


\node  (L) [below = 4.75cm of C,draw=teal!50,align=center,rectangle,rounded corners,minimum height=6cm,minimum width=15cm, text width=5.5cm]  {\tiny};

\node (L1) [below = 5.5cm of C,xshift=3cm,align=center, text width=5 cm] { \textbf{\scriptsize Quantum proposal approach}};

\node (L2) [align=center, text width=3 cm,draw=teal!50,rounded corners] at (-5.5,-1.35) { \textbf{\scriptsize Quantum part}};

\node (L3) [below = 5.5cm of C,xshift=-4cm,align=center, text width=5 cm] { \textbf{\scriptsize Quantum Euclidean distance}};

\node(LSwap) [align=center,below = 6cm of C,xshift=-4cm,scale=0.75]  {

\Qcircuit @C=1em @R=1em {
\lstick{\ket{0}}& \gate{H} & \qw   & \qw & \ctrl{2} & \gate{H} & \meter \\
\lstick{q_{1}} & \qw & \gate{\psi}  & \qw & \qswap & \qw & \qw \\
\lstick{q_{2}} & \qw & \multigate{2}{\phi} &  \qw & \qswap & \qw & \qw \\
\lstick{q_{3}} & \qw & \ghost{\phi} & \qw & \qw & \qw & \qw\\
\lstick{q_{4}} & \qw & \ghost{\phi} & \qw & \qw & \qw & \qw\\
\lstick{C} & \qw & \qw &\qw &\qw & \qw & \qw \cwx[-5]} 

};

\node (L4) [below = 9.75cm of C,xshift=-4cm,draw=violet!50,align=center,rectangle,rounded corners, text width=5 cm] { \textbf{\scriptsize qBraid framework}};

\node(Proposal) [align=center,below = 6cm of C,xshift=3.35cm,scale=0.5]{
\Qcircuit @C=.01em @R=.75em {
\lstick{\ket{a^1}}& \multigate{2}{Img_1} & \qw & \qw & \ctrl{6} & \qw & \qw & \qw & \qw & \qw & \qw & \qw & \qw & \qw & \qw & \qw\\
\lstick{\ket{\vdots}} & \ghost{Img_1} & \qw & \qw & \qw & \ctrl{5} & \qw & \qw & \qw & \qw & \qw & \qw & \qw & \qw & \qw & \qw\\
\lstick{\ket{a^n}} & \ghost{Img_1} & \qw & \qw & \qw & \qw & \ctrl{4} & \qw & \qw & \qw & \qw & \qw & \qw & \qw & \qw & \qw\\
\lstick{\ket{b^1}} & \multigate{2}{Img_2} & \qw & \qw & \qw & \qw & \qw & \ctrl{3} & \qw & \qw & \qw & \qw & \qw & \qw & \qw & \qw\\
\lstick{\ket{\vdots}} & \ghost{Img_2} & \qw & \qw & \qw & \qw & \qw & \qw & \ctrl{2} & \qw & \qw & \qw & \qw & \qw & \qw & \qw\\
\lstick{\ket{b^n}} & \ghost{Img_2} & \qw & \qw & \qw & \qw & \qw & \qw & \qw & \ctrl{1}  & \qw & \qw & \qw & \qw & \qw & \qw\\
\lstick{\ket{0}} & \qw & \qw & \multigate{2}{QFT} &  \multigate{2}{QDA} & \multigate{2}{QDA} & \multigate{2}{QDA} & \multigate{2}{QDA} & \multigate{2}{QDA} & \multigate{2}{QDA} & \qw & \qw & \qw & \multigate{2}{inv QFT} & \qw & \meter\\
\lstick{\ket{\vdots}} & \qw & \qw & \ghost{QFT} & \ghost{QDA} & \ghost{QDA} & \ghost{QDA} & \ghost{QDA} & \ghost{QDA} & \ghost{QDA} & \qw & \qw & \qw & \ghost{inv QFT} & \qw & \meter\\
\lstick{\ket{n}} & \qw & \qw & \ghost{QFT} & \ghost{QDA} & \ghost{QDA} & \ghost{QDA} & \ghost{QDA} & \ghost{QDA} & \ghost{QDA} & \qw & \qw & \qw & \ghost{inv QFT}
& \qw & \meter}

};

\node (L4) [below = 9.75cm of C,xshift=3cm,draw=black,align=center,rectangle,rounded corners, text width=5 cm] { \textbf{\scriptsize IBM framework}};

\node  (qbraid) [below = 5.5cm of C,xshift=-4cm,draw=violet!50,align=center,rectangle,rounded corners,minimum height=5cm,minimum width=6cm]  {\tiny};

\node  (ibm) [below = 5.5cm of C,xshift=3.1cm,draw=black,align=center,rectangle,rounded corners,minimum height=5cm,minimum width=7.5cm]  {\tiny};




\node  (P) [below = .5cm of L,draw=black,align=center,rectangle,draw=black,rounded corners,minimum height=1.5cm,minimum width=7cm, text width=7cm] {\textbf{\scriptsize Noise models with 1-qubit or 2-qubit gates}};

\node (AWS) [below = .5cm of P,draw=violet!50,align=center,rectangle,rounded corners,minimum height=1.5cm,minimum width=7cm, text width=7cm] {\includegraphics[scale=.035]{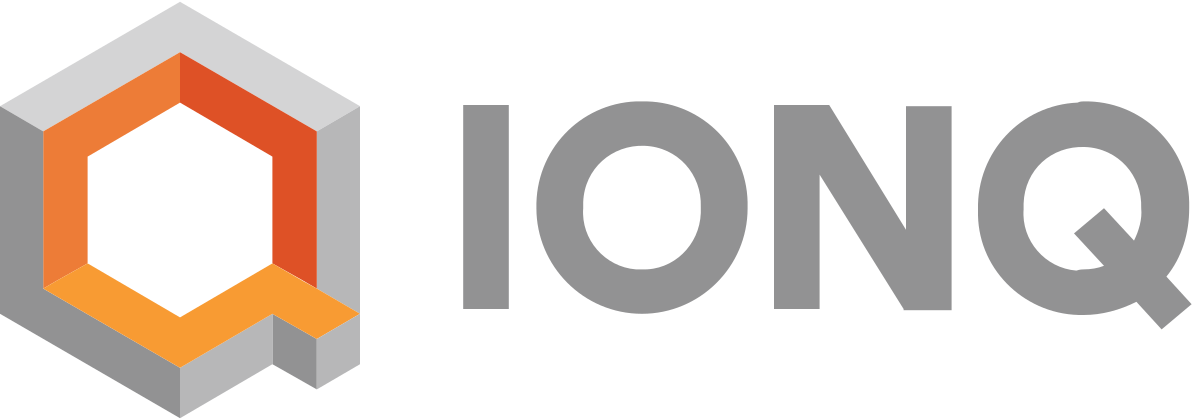}\includegraphics[scale=.35]{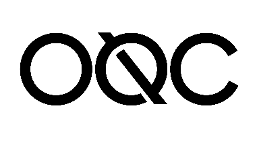} \includegraphics[scale=.035]{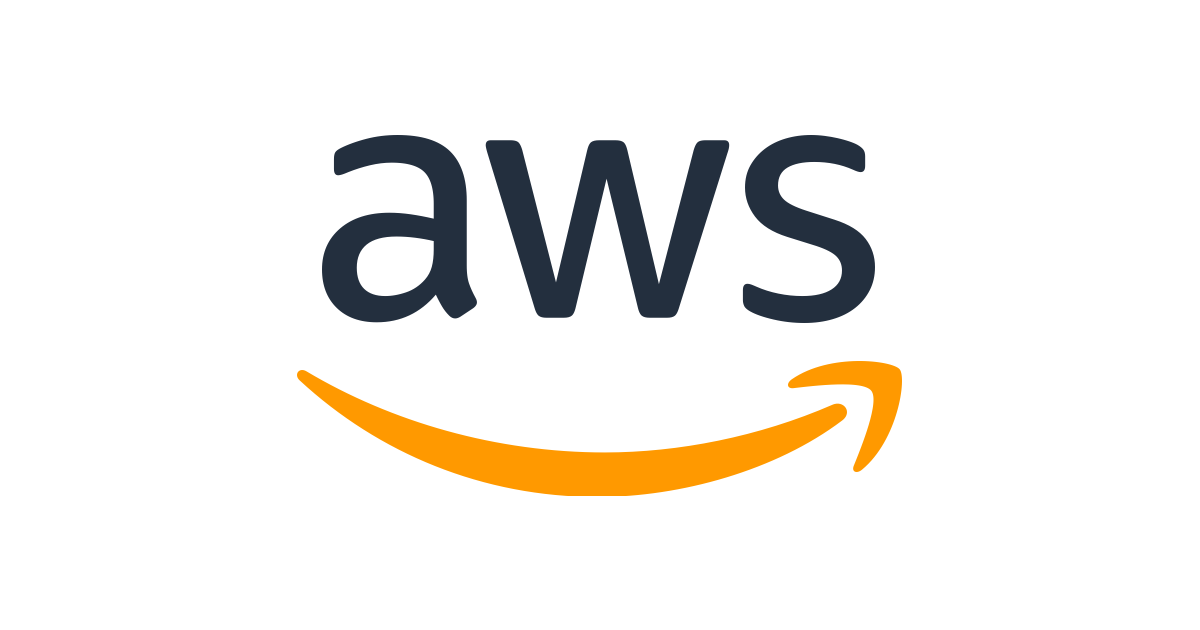} \includegraphics[scale=.035]{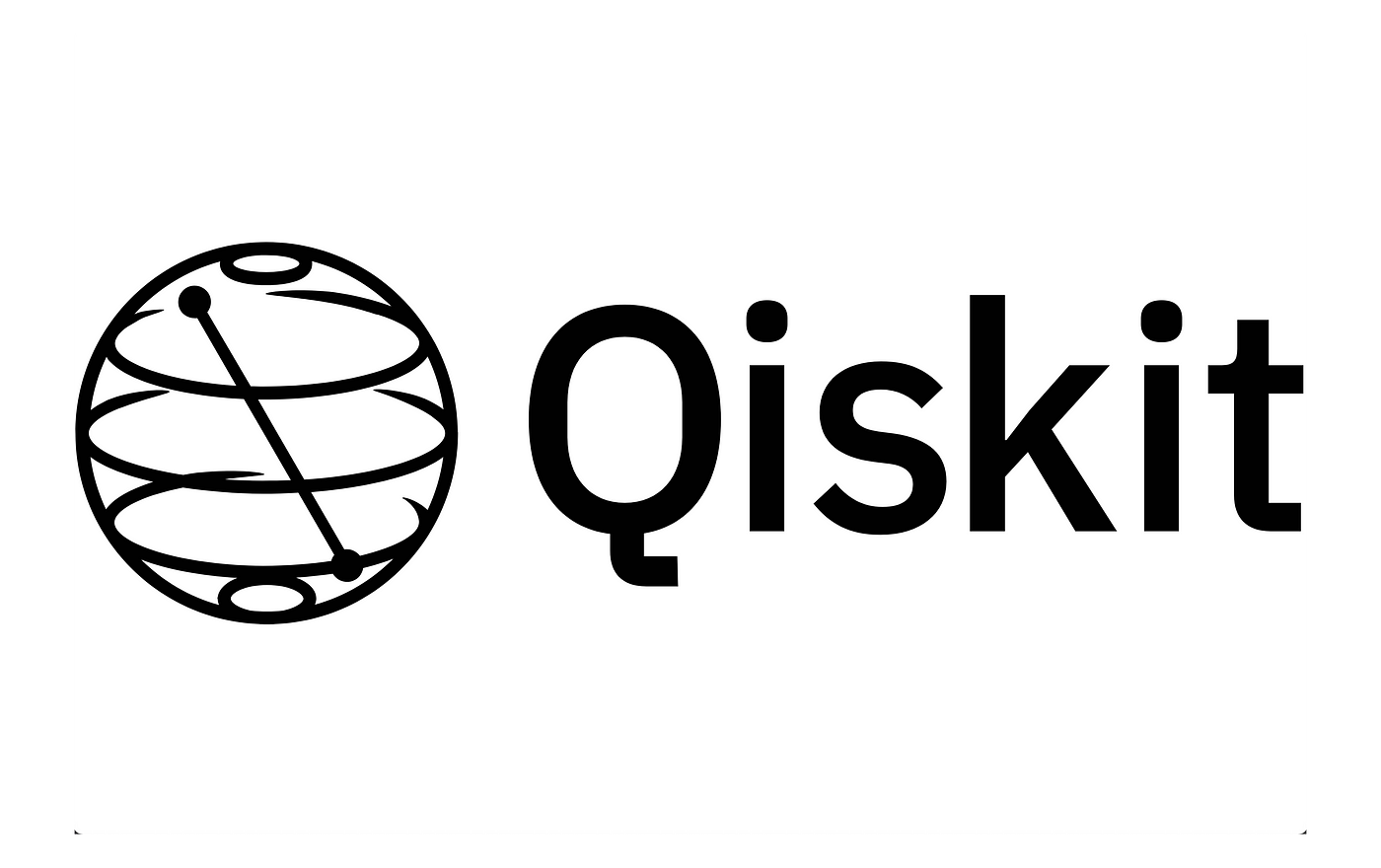}};

\node  (R) [below = .5cm of AWS,align=center,rectangle,draw=cyan,rounded corners,minimum height=1.5cm,minimum width=5cm, text width=5cm]  {\textbf{\scriptsize Results from quantum circuits}};


\draw [->][color=black,dashed,line width=1.5pt] (A2) -- (J);
\draw [->][color=black,dashed,line width=1.5pt] (J) -- (L);
\draw [->][color=black,dashed,line width=1.5pt] (L) -- (P);
\draw [->][color=black,dashed,line width=1.5pt] (P) -- (AWS);
\draw [->][color=black,dashed,line width=1.5pt] (AWS) -- (R);

\end{tikzpicture}
}
    \caption{Diagram of the proposal methodology for a quantum block-matching algorithm using dissimilarity measure.}
    \label{fig:general_diagram}
    
\end{figure}

The proposal work is introduced, and the general diagram of the present methodology is shown in   \autoref{fig:general_diagram}.

\subsection{Classical approach}
\begin{figure*}[h] 
\centering
\subfloat[]{\includegraphics[width=.2\linewidth]{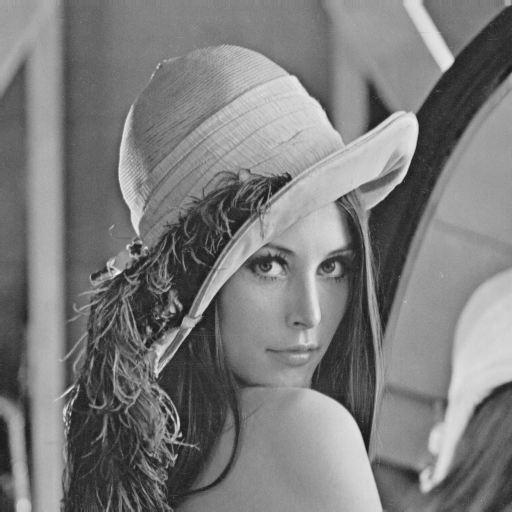}}
\hfil
 \subfloat[]{\includegraphics[width=.2\linewidth]{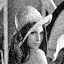}}
  \caption{ (a) The original image. (b) The original image with Gaussian noise $\mu=0$ and $\sigma = 20$ and 4-bit grayscale.}
  \label{fig:Images_grayscale}
\end{figure*}

In this first part, it is essential to resize the tested images in order to obtain experimental results implementing quantum circuits. So, the original images are contaminated with Gaussian noise $I_{r} = I_{o} + G$, where $I_{r}$ is the image with noise, $I_{o}$ is the original image and $G$ is a normally distributed random variable of mean $\mu$ and variance $\sigma^{2}$. In the present work $\mu = 0$ and $\sigma^{2} = 20$ \cite{02QIEEE}. after this, since the original images are $512 \times 512$ pixels, size reduction is applied to obtain an image with $64 \times 64$ pixels. For the last stage in image processing, a grayscale palette reduction is applied from 8-bit grayscale to 4-bit grayscale \autoref{fig:Images_grayscale}.

The resulting image from \autoref{fig:Images_grayscale} can be smoothed in a spatial domain transform or a low pass filter. Such transformation can introduce some robustness in a noisy environment. In order to reduce the complexity of some operations, a hierarchical search algorithm is implemented in comparison to the full search algorithm \cite{01Libro}. With this proposal, the complexity can be reduced from O($n^{2}K$) to O($(\frac{n}{2})^{2}K$), where $K$ is the search area and $n$ is the size of the window for each patch image. Once the search algorithms are introduced, the next step is to define the experiment parameters. In the present work, $k = 10$ and $n=8$ are chosen; considering the hierarchical search, the size of the vectors is reduced from $n=8$ to $n=4$.

\subsection{Quantum approach}

The standard dissimilarity measure is obtained between two image blocks. Firstly, the image blocks are flattened in vectors of size $r_x$. Secondly, to implement the measure of Euclidean distance $D$, the QFT or the Quantum swap test is implemented; it can be formulated as \cite{Qiskit}:

\begin{small}
\begin{equation}\label{eqn:4}
\begin{aligned}
&D\left(Img_{1}, Img_{2}\right)= \\
&\sum_{i=0}^{r_{x}-1}\begin{array}{l}
\left[Img_{1}(x+i)-Img_{2}\left(x+o_{x}+i\right)\right]^{2}
\end{array},
\end{aligned}
\end{equation}
\end{small}

\subsubsection{QFT approach}
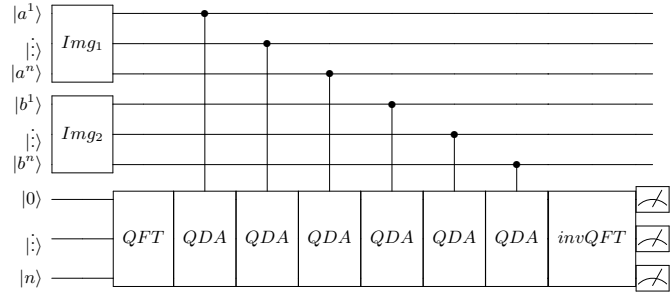
\begin{figure}[h]
    \centering
    \begin{tikzpicture}[]
 \node[scale=0.75] (qcircuit){\Qcircuit @C=.01em @R=.75em {
\lstick{\ket{a^1}}& \multigate{2}{Img_1} & \qw & \qw & \ctrl{6} & \qw & \qw & \qw & \qw & \qw & \qw & \qw & \qw & \qw & \qw & \qw\\
\lstick{\ket{\vdots}} & \ghost{Img_1} & \qw & \qw & \qw & \ctrl{5} & \qw & \qw & \qw & \qw & \qw & \qw & \qw & \qw & \qw & \qw\\
\lstick{\ket{a^n}} & \ghost{Img_1} & \qw & \qw & \qw & \qw & \ctrl{4} & \qw & \qw & \qw & \qw & \qw & \qw & \qw & \qw & \qw\\
\lstick{\ket{b^1}} & \multigate{2}{Img_2} & \qw & \qw & \qw & \qw & \qw & \ctrl{3} & \qw & \qw & \qw & \qw & \qw & \qw & \qw & \qw\\
\lstick{\ket{\vdots}} & \ghost{Img_2} & \qw & \qw & \qw & \qw & \qw & \qw & \ctrl{2} & \qw & \qw & \qw & \qw & \qw & \qw & \qw\\
\lstick{\ket{b^n}} & \ghost{Img_2} & \qw & \qw & \qw & \qw & \qw & \qw & \qw & \ctrl{1}  & \qw & \qw & \qw & \qw & \qw & \qw\\
\lstick{\ket{0}} & \qw & \qw & \multigate{2}{QFT} &  \multigate{2}{QDA} & \multigate{2}{QDA} & \multigate{2}{QDA} & \multigate{2}{QDA} & \multigate{2}{QDA} & \multigate{2}{QDA} & \qw & \qw & \qw & \multigate{2}{inv QFT} & \qw & \meter\\
\lstick{\ket{\vdots}} & \qw & \qw & \ghost{QFT} & \ghost{QDA} & \ghost{QDA} & \ghost{QDA} & \ghost{QDA} & \ghost{QDA} & \ghost{QDA} & \qw & \qw & \qw & \ghost{inv QFT} & \qw & \meter\\
\lstick{\ket{n}} & \qw & \qw & \ghost{QFT} & \ghost{QDA} & \ghost{QDA} & \ghost{QDA} & \ghost{QDA} & \ghost{QDA} & \ghost{QDA} & \qw & \qw & \qw & \ghost{inv QFT}
& \qw & \meter}};
\end{tikzpicture}
    \caption{Quantum Circuit for the QFT approach.}
    \label{fig:QFT}
\end{figure}

In order to apply the Quantum Draper Adder, first, the qubits where the sum is stored are initialized using the Hadamard gate ($H$); after this, the rotation of each pixel value is coded. For save numbers of rotations, the Quantum Phase Estimator is introduced to encode the pixel value of each vector $\vec{Img_1}^{\,}$ and $\vec{Img_2}^{\,}$. After this, the inverse quantum Fourier transform is implemented. Finally, a measurement is executed in the qubits, where the sum is stored. The quantum circuit is shown in \autoref{fig:QFT}.

\subsubsection{Quantum Swap Test approach}

\begin{figure}[h]
    \centering
    \includegraphics[width=.70\linewidth]{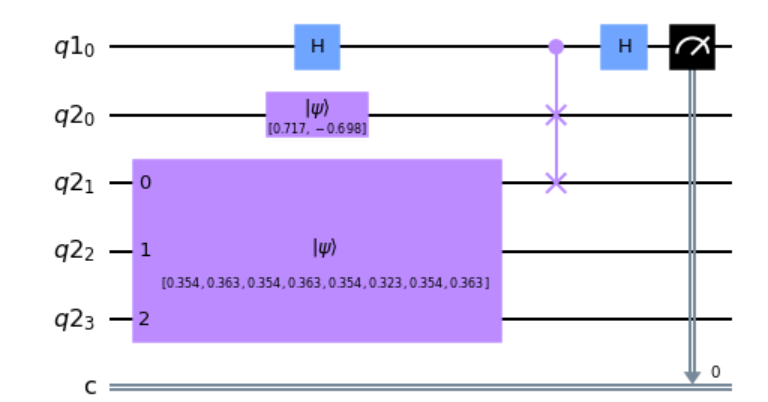}
    \caption{Quantum Euclidean Distance using Swap Test approach with the vectors $\vec{Img_1}^{\,} = [ 9, 9, 9, 9 ] $ and $\vec{Img_2}^{\,} = [ 9, 9, 8, 9 ] $.}
    \label{fig:QST}
\end{figure}

\noindent The advantage of this approach is the number of qubits. In performing the quantum circuit Swap Test, $n = log_{2}(M) + 1$ qubits are implemented. From the \autoref{eqn:2} and \autoref{eqn:3}, the dissimilarity measure can be executed. So $\ket{\psi}$ and $\ket{\phi}$ are obtained since $A = \vec{Img_1}^{\,}$ and $B= \vec{Img_2}^{\,}$. the quantum circuit is shown in the next \autoref{fig:QST}.

\subsection{Quantum noise models}

A method to analyze the behavior of quantum circuits before running it on a QPU is to create a noise model that considers the noise in the real QPU. The main errors in a QPU are decoherence and gate errors. We focused on a noise model with different gate errors - fidelities of the two qubits gates. Decreasing the fidelity will affect the results considerably. Finding a reasonable fidelity in the simulations will give us an estimate of the resources we will need for a perfect outcome of the circuits in real QPU.

\section{Results} \label{Sect:4}
The experiments were implemented with Qiskit \cite{Qiskit} using the \texttt{qBraid} environment \cite{qBraid} and with AWS braket in the case of Hardware jobs with IonQ and Oxford Quantum Circuits (OQC).

\subsection{Results for QFT approach}

\begin{figure}[h]
    \centering
    \includegraphics[width=.8\linewidth]{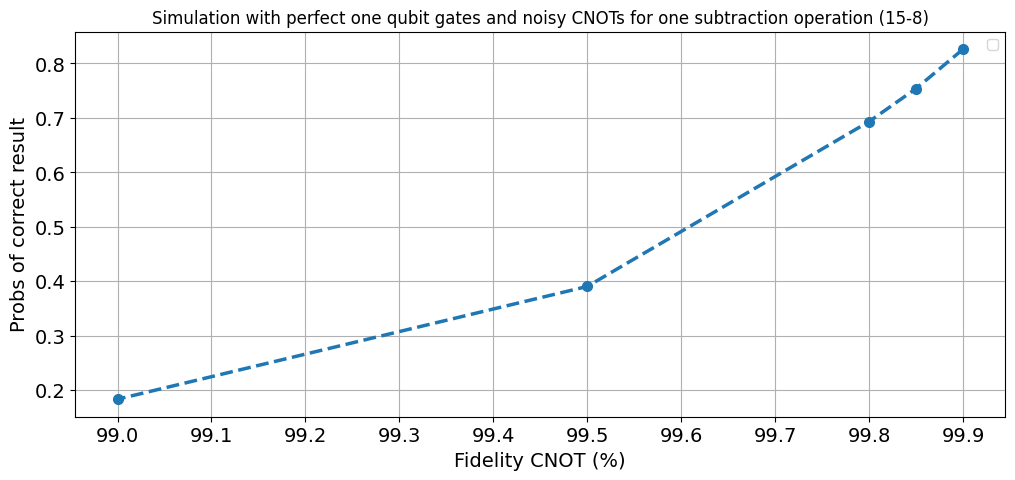} 
    \vspace{1ex}
  \caption{Noisy simulation with different CNOT gate fidelities using the QFT approach. }
  \label{fig:results_sim_qft}
\end{figure}

The results using the QFT approach are ideal in the perfect simulation. Nevertheless, many qubits are needed since the quantum circuits have a considerable depth. Therefore, this approach is unreasonable for the current NISQ devices. To showcase the disadvantage of this approach, a simple subtraction operation of two integers, as shown in the \autoref{fig:results_sim_qft} is simulated. In the worst case, at least 12 qubits with more than 200 CNOT gates will be used. In implementing the BM problem, each vector represents sections of an image with multiple subtractions needed to compute the Euclidean distance. Despite this approach having many quantum resources, the result will always be accurate in a fault-tolerant device. The error increases rapidly once a slight noise into the CNOT gates is introduced. Currently, improvements to the resources needed for this approach are being made. 

Given the high number of quantum resources needed for the QFT approach, the Swap Test, a more suitable algorithm for NISQ devices, is explored.

\subsection{Results for Swap Tets approach}

The results with the Swap Test approach were done with 17 pairs of vectors of 4 dimensions. Based on the classical part similar to those representing the different sections of images for the BM problem.

\begin{figure}[h]
\centering
\includegraphics[width=.9\linewidth]{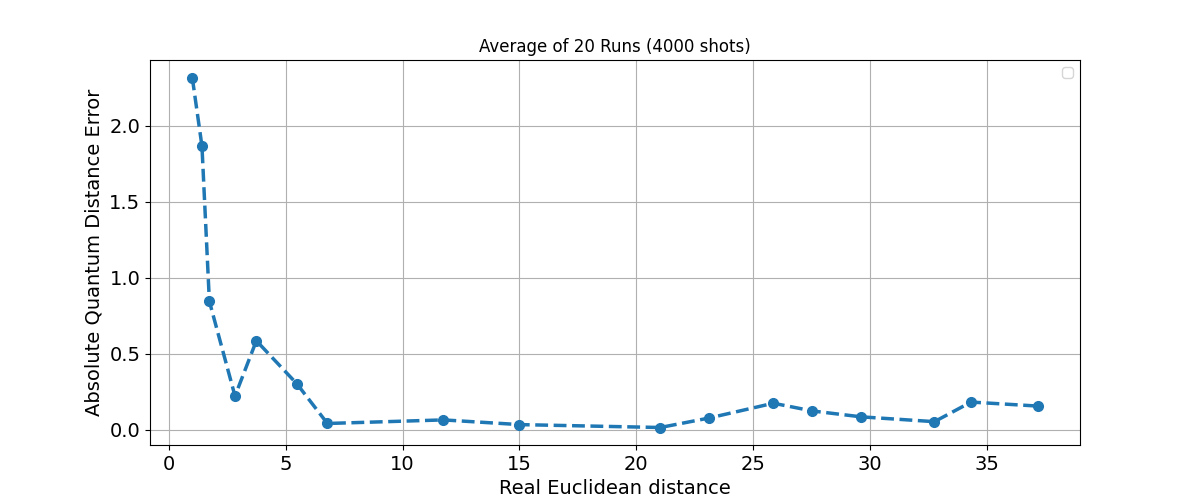}
  \caption{Noiseless simulation of the computation of the euclidean distance using the Swap Test for 17 pairs of vectors. With an average of 20 runs and 4000 shots each. }
  \label{fig:results_sim_swap}
\end{figure}

\begin{figure}[h]
    \centering
    \includegraphics[width=.9\linewidth]{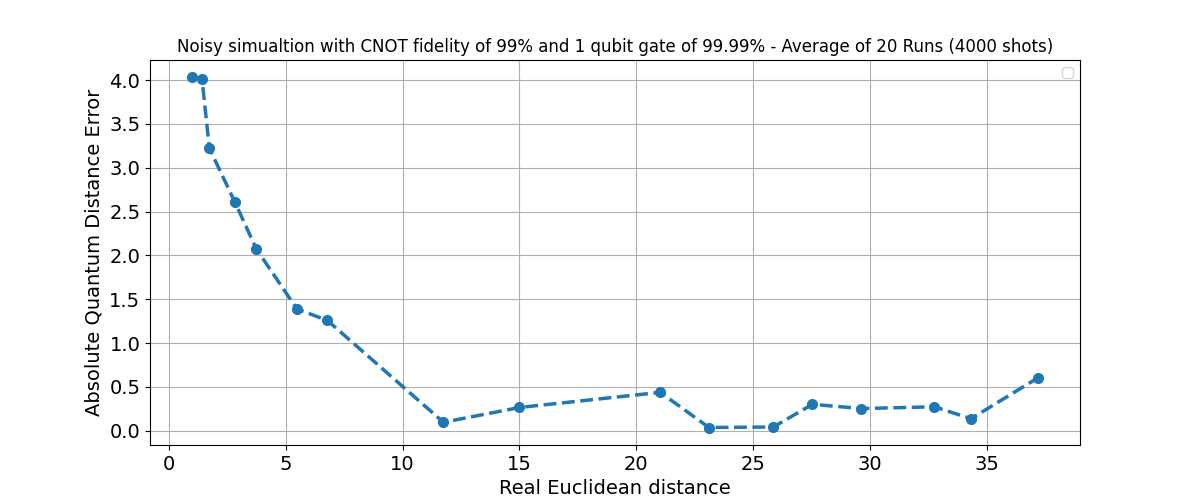} 
  \caption{Noisy simulation with a CNOT fidelity of 99 \%  and one qubit gate of 99.99 \% of the Swap Test for 17 pairs of vectors. With an average of 20 runs and 4000 shots each.}
  \label{fig:noisy_results_sim_swap}
\end{figure}

The \autoref{fig:results_sim_swap} shows results with noiseless simulations of 20 runs with 4000 shots each for the different vectors. In \autoref{fig:noisy_results_sim_swap}, a noise simulation results with depolarising errors in the CNOTs (with 99 \% gate fidelity), and one qubit gate (with 99.99 \% gate fidelity) is shown. For both cases, the results are accurate when the distance is more significant, starting on the order of 5 units in the case of perfect simulation and 10 in the noisy model. \\

\begin{figure}[h]
    \centering
    \includegraphics[width=0.8\linewidth]{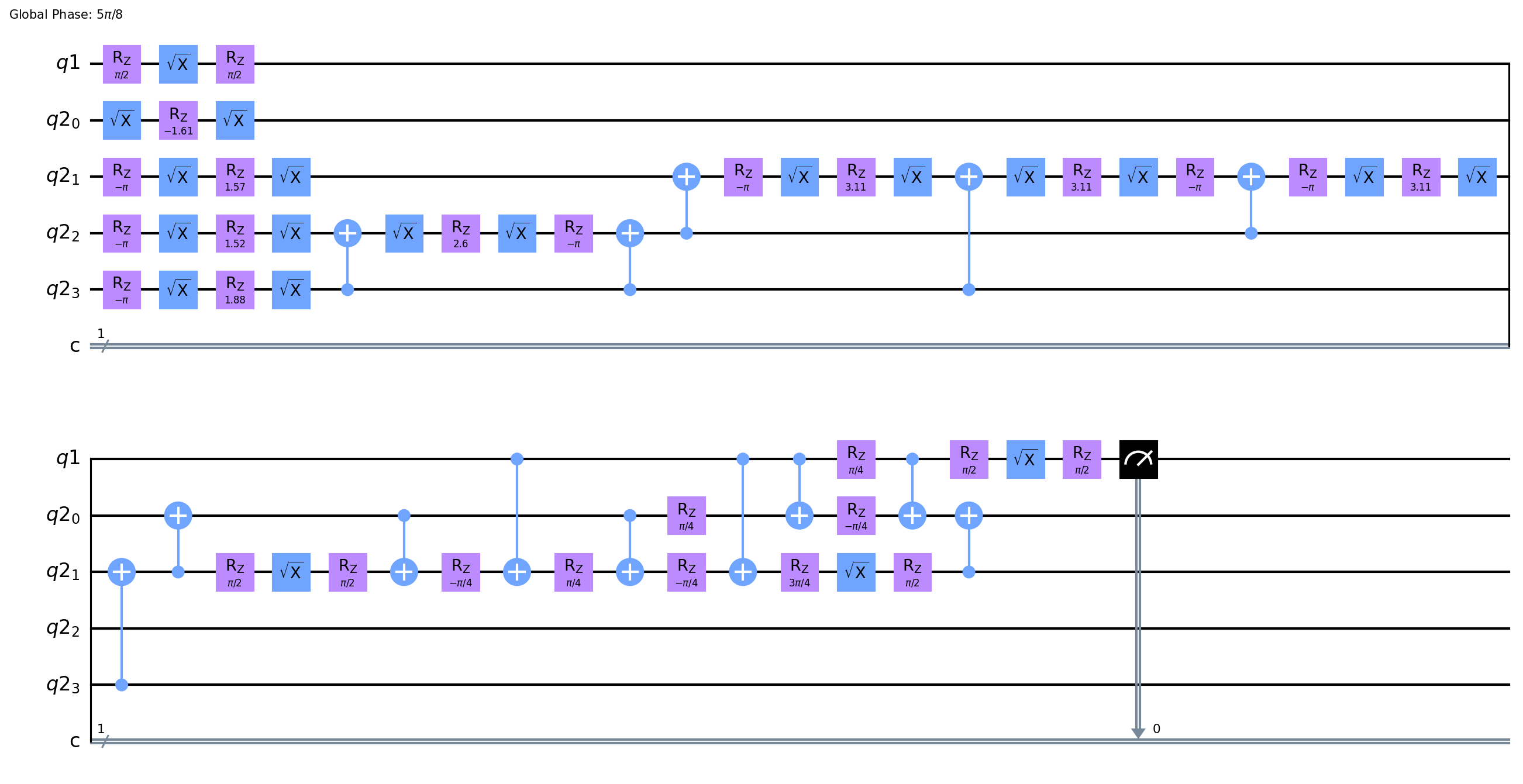} 
  \caption{Transpiled circuit for the Swap Test with the gate basis $[cx, id, rz, sx]$ }
  \label{fig:transpiled_swap}
\end{figure}

The error in the small distance area is attributed to the encoding in the angle of rotations; when the two vectors are close to each other, the angle will become small enough to create errors. As seen in the transpiler circuit shown in \autoref{fig:transpiled_swap}, the swap test involves only fewer CNOT gates than the QFT approach.\\

Given the applicability of the Swap test in current NISQ devices, the experiments for different pairs of vectors in several QPUs using the \texttt{qBraid} environment are executed. The proposal quantum-classical hybrid block-matching algorithm with the image vectors $\vec{Img_1}^{\,} = [ 9, 9, 9, 9 ] $ and $\vec{Img_2}^{\,} = [ 9, 9, 8, 9 ] $ in the following QPU's: IonQ Harmony device, Oxford Quantum Circuits (OQC) and IonQ Aria device and Ibmq-Belem were executed. The experimental results are shown in the \autoref{tab2} with the number of qubits and shots used in each case.

\begin{table}[h]
    \begin{center}
\caption{Comparison between the Quantum Euclidean Distance (QED) and the Classical Euclidean Distance (CED) using vectors $\vec{Img_1}^{\,}$ and $\vec{Img_2}^{\,}$ .The proposal quantum-classical hybrid block-matching algorithm was executed in IonQ Harmony, OQC, IonQ Aria, and Ibmq-Belem.}\label{tab2}
\begin{tabular}{|l|l|l|l|l|}
\hline
Provider &  Qubits & Shots & QED & CED\\
\hline
Ibmq-Belem &  5 &1000 & 6.55 & 1\\
OQC &  8 & 1000& 15.887 	 & 1\\
IonQ Harmony &  11 &1000 & 7.452 & 1\\
IonQ Aria &  25 & 1000& 10.173 & 1\\
Ibmq-nairobi &  7 & 4000 & 5.54 & 1\\
Ibmq-nairobi &  7 & 4000& 16.74 & 11.747\\
Ibmq-nairobi &  7 & 4000 & 26.82 & 23.11\\
\hline
\end{tabular}
\end{center}
\end{table}

\section{Conclusions} \label{Sect:5}

This work presented a proof of concept of the standard dissimilarity measure using the Euclidean distance in a quantum computing approach. Also, the hierarchical search has been implemented to reduce the size of the vectors and encode the pixel value between two image blocks for the quantum approach. In this case, only the dissimilarity of a set of 17 pairs of vectors was obtained. However, the approach can be scaled to a complete image showing an application of quantum computing in image processing. In summary, based on the experiments executed in quantum processing units, we found that when more dissimilarity is considered, the algorithms based on the swap test obtain results very close to the classical Euclidean distance, assuming a fidelity in quantum computers of 99\%, with promising results for future applications Within the noisy intermediate-scale quantum era.

The depth of the quantum circuit based on the Quantum Fourier Transform approach using less quantum resources can be made as a future work. Even though, to obtain results in quantum processing units, a quantum error correction approach can be implemented. In addition, we plan to use this work in a sequence of digital videos.      

\section*{Acknowledgments}

This work is supported by the Consejo Nacional de Humanidades, Ciencias y Tecnologías (CONAHCYT), Instituto Politécnico Nacional (IPN) and Quantum Open Source Foundation (QOSF). Also this work was sponsored by qBraid, a cloud-based platform for quantum computing. It provides software tools for researchers and developers in quantum as well as access to quantum and classical hardware.


%
%

%
%
%
%

\bibliographystyle{splncs04}
\bibliography{samplepaper}
\end{document}